\documentclass[showpacs,10pt,twocolumn,prl]{revtex4-1}

\usepackage{amsmath}
\usepackage{amssymb}
\usepackage{graphicx}
\usepackage{amssymb}
\usepackage{graphics}
\usepackage{epsfig}
\usepackage{CJK}
\usepackage{color}
\usepackage{soul}

\begin{document}

\begin{CJK*}{GBK}{Song}
\title{Enhanced Superconductivity and Electron Correlations in Intercalated ZrTe$_3$}
\author{Yu Liu,$^{1,\S,\P}$ Xiao Tong,$^{2}$ V. N. Ivanovski,$^{3}$ Zhixiang Hu,$^{1,4}$ Denis Leshchev,$^{5}$ Xiangde Zhu,$^{1,*}$ Hechang Lei,$^{1,\dag}$ Eli Stavitski,$^{5}$ Klaus Attenkofer,$^{5,\ddag}$ V. Koteski,$^{3}$ and C. Petrovic$^{1,4,\|}$}
\affiliation{$^{1}$Condensed Matter Physics and Materials Science Department, Brookhaven National Laboratory, Upton, New York 11973, USA\\
$^{2}$Center of Functional Nanomaterials, Brookhaven National Laboratory, Upton, New York 11973, USA\\
$^{3}$Department of Nuclear and Plasma Physics, Vinca Institute of Nuclear Sciences - National Institute of the Republic of Serbia, University of Belgrade, Belgrade 11001, Serbia\\
$^{4}$Materials Science and Chemical Engineering Department, Stony Brook University, Stony Brook, New York 11790, USA\\
$^{5}$National Synchrotron Light Source II, Brookhaven National Laboratory, Upton, New York 11973, USA}
\date{\today}

\begin{abstract}
  Charge density waves (CDWs) with superconductivity, competing Fermi surface instabilities and collective orders, have captured much interest in two-dimensional van der Waals (vdW) materials. Understanding of CDW suppression mechanism, its connection to emerging superconducting state and electronic correlations provides opportunities for engineering the electronic properties of vdW heterostructures and thin film devices. Using combination of the thermal transport, X-ray photoemission spectroscopy, Raman measurements, and first-principle calculations, we observe an increase in electronic correlations of the conducting states as CDW is suppressed in ZrTe$_3$ with 5\% Cu and Ni intercalation in the vdW gap. As superconductivity emerges, intercalation brings decoupling of quasi-one-dimensional conduction electrons with phonons as a consequence of intercalation-induced lattice expansion but also a drastic increase in Zr$^{2+}$ at the expense of Zr$^{4+}$ metal atoms. These observation demonstrate the potential of atomic intercalates in vdW gap for ground state tuning but also illustrate the crucial role of Zr metal valence in formation of collective electronic orders.
\end{abstract}
\maketitle
\end{CJK*}

\section{INTRODUCTION}

Electronic correlations associated with charge density wave (CDW) and superconductivity (SC), both the Fermi surface (FS) instabilities and low-temperature collective orders in solids, attract considerable attention \cite{Gruner,CastroNeto,Schmitt,Kivelson}. In particular, layered van der Waals (vdW) transition metal chalcogenides raised interest due to possible quantum critical phenomena \cite{BarathH,JoeY}; yet due to relatively easy exfoliation, heteroepitaxial growth, and device fabrication of vdW crystals they can be used to investigate confinement of correlated electronic wave functions involved in CDW and SC and also to engineer new quantum states \cite{GeremewA,ZhangZ,ConstantCB,ZhouX,HanGH}.

ZrTe$_3$ provides a good platform due to low-dimensional crystal structure with $P2_1/m$ unit cell symmetry [Fig. 1(a)]: quasi-two-dimensional (2D) ZrTe$_3$ prisms aligned along the $c$-axis are separated by the vdW gap with quasi-one-dimensional (1D) ZrTe$_6$ chains running along the $b$-axis; in addition there are Te2-Te3 chains along the $a$-axis \cite{Bra,Fur,Wieting}. ZrTe$_3$ shows not only CDW ($T_{\textrm{CDW}}$ $\sim$ 63 K) with Peierls modulation $\emph{\textbf{q}}\approx (\frac{1}{14}, 0, \frac{1}{3})$ but also a filamentary SC ($T_\textrm{c} \sim$ 2 K) that arises on cooling from a quasi-2D electronic conduction (resistivity $\rho_a\approx\rho_b\sim\frac{\rho_c}{10}$) \cite{Eaglesham,Takahash,Nakajima,Yamaya}. A local-pairs-induced SC mechanism was proposed \cite{Tsu}. Band structure calculation and angular resolved photoemission (ARPES) measurements revealed that the FS consists mainly of a three-dimensional (3D) FS sheet centered at the Brillouin zone (BZ) center and the quasi-1D FS sheets parallel to the inclination of the BZ boundary \cite{Stowe,Fesler,Star,Hoesch,Lyu}. A pseudogap feature and CDW fluctuations persist far above $T_{\textrm{CDW}}$ \cite{Per,Yokoya,Chinotti}. Kohn anomaly associated with a soft phonon mode was identified \cite{Hoesch1}, while the Raman experiment also emphasized the important role of electron-phonon (e-p) coupling at ambient or high pressure \cite{Hu,Gleason}. Additionally, mixed valence of Zr in ZrTe$_3$ nanoribbons was reported \cite{Yu}. Valence segregation with SC in transition metal oxides with different electron correlation strength can be connected to SC mechanism and the e-p coupling \cite{GoodenoughJB,VarmaCM,RiceTM,KimM}.

Bulk SC can be induced in ZrTe$_3$ by applying physical pressure, intercalation, doping, and disorder \cite{Yomo,Hoesch2,Gu,HC,XD1,Yadav,Zhu,XD2,Cui,Ganose,Yan,Yue}. A pressure-induced re-entrant SC in ZrTe$_3$ implies a possible unconventional SC mechanism \cite{Yomo}, while the ultra-low-temperature thermal conductivity indicates multiple nodeless gaps in ZrTe$_{3-x}$Se$_x$ \cite{Cui}. Light intercalation of $\sim$ 5\% Cu and Ni can induce bulk SC with $T_\textrm{c} \sim$ 3.8 and 3.1 K in Cu$_{0.05}$ZrTe$_3$ and Ni$_{0.05}$ZrTe$_3$, respectively, that coexists with the CDW order \cite{HC,XD1}. The CDW-related resistivity anomaly for the current flow along the $a$-axis shifts to lower temperatures with reduced amplitudes, however, the optical absorption spectra show an increase of the CDW gap size in (Cu,Ni)$_{0.05}$ZrTe$_3$ \cite{Mirri}.

\begin{figure}
\centerline{\includegraphics[scale=1]{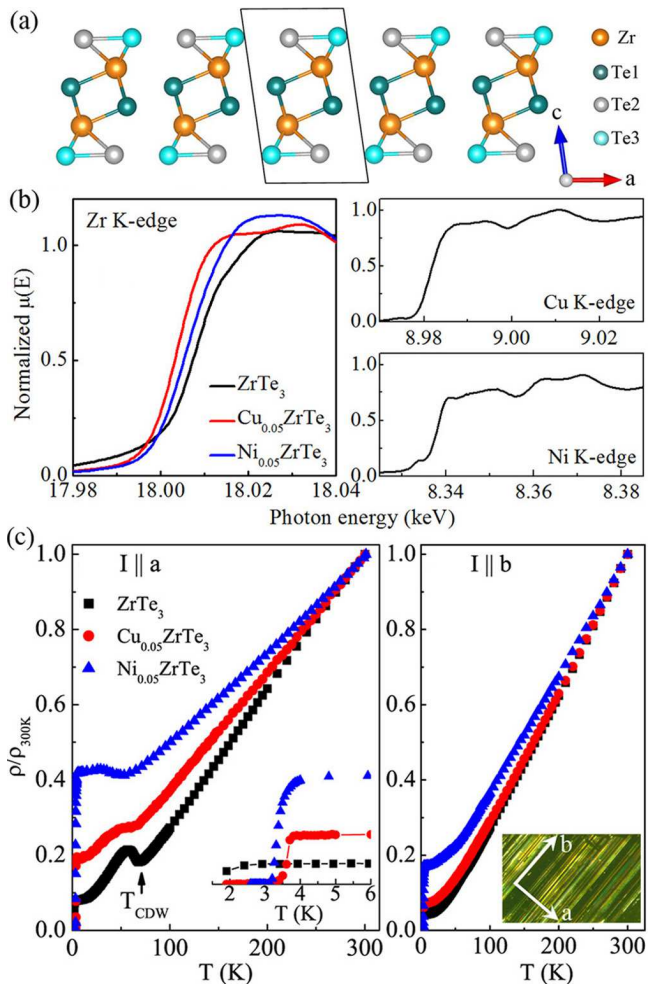}}
\caption{(Color online) (a) Crystal structure of ZrTe$_3$ with quasi-1D ZrTe$_6$ chains along the $b$-axis and the Te2/Te3 rectangular network layer viewed from the $c$-axis. (b) Normalized Zr (left), Cu and Ni (right) $K$-edge XANES spectra at room temperature. (c) Temperature dependence of the normalized resistivity $\rho$(T)/$\rho$(300 K) for ZrTe$_3$ and (Cu,Ni)$_{0.05}$ZrTe$_3$ with the current flow along the $a$-axis (left) and the $b$-axis (right). Insets show the low-temperature superconducting transitions and current flow direction in crystals.}
\label{XRD}
\end{figure}

Thermopower measurement is suitable to characterize the nature and sign of carries as well as the correlation strength in superconductors whereas X-ray photoemission spectroscopy (XPS) and Raman measurements are good probes of the valence state and phonon vibrations in transition metal compounds \cite{Collignon,Kefeng,Behnia,Pourret,Demeter}. Herein we examine the electron correlation strength and the Zr valence in superconducting (Cu,Ni)$_{0.05}$ZrTe$_3$ single crystals. Thermal transport and Raman measurements show a suppressed CDW in (Cu,Ni)$_{0.05}$ZrTe$_3$ when compared to ZrTe$_3$. An obvious anomaly around $T_\textrm{CDW}$ in thermal conductivity $\kappa$(T) was observed in ZrTe$_3$, indicating strong e-p coupling. The anomaly is suppressed with weaker amplitudes in (Cu,Ni)$_{0.05}$ZrTe$_3$. A clear slope change in thermopower $S$(T) corresponds to a FS reconstruction across the CDW transition. Both the ratio of $T_\textrm{c}$ to the Fermi temperature $T_\textrm{F}$ and the effective mass $m^*$ rise in (Cu,Ni)$_{0.05}$ZrTe$_3$ when compared to ZrTe$_{2.96}$Se$_{0.04}$ imply increase in electronic correlations of the bulk conducting states due to contribution of Cu and Ni, in agreement with the first-principle calculations. Whereas intercalation in the vdW gap stabilizes bulk SC at the expense of CDW order, it coincides with drastic reduction of Zr mixed valence and the expansion of unit cell along the $a$- and $c$-directions as well as a rapid increase in atomic concentration ratio of Zr$^{2+}$ to Zr$^{4+}$ induced by Cu,Ni intercalation. This affects the Zr 4$d$-derived electronic states near the $\Gamma$ point in the Brillouen zone below the Fermi level.

\section{EXPERIMENTAL DETAILS}

High quality single crystals of ZrTe$_3$ with $\sim$ 5\% Cu,Ni intercalation were fabricated by chemical vapor transport method \cite{HC,XD1}. X-ray absorption near edge spectroscopy measurement was performed at 8-ID beamline of the National Synchrotron Light Source II at Brookhaven National Laboratory in the fluorescence mode. Thermal transport was measured in quantum design PPMS-9 with standard four-probe technique. Thermopower was measured by using one-heater-two-thermometer setup with hooked copper leads using silver paint contact directly on crystals. Continuous measuring mode was adopted for thermopower measurement with the maximum heater power and period set as 50 mW and 1430 s along with the maximum temperature rise of 3$\%$. The relative error in our measurement for thermopower was below 5\% based on Ni standard measured under identical conditions. Sample dimensions were measured by an optical microscope Nikon SMZ-800 with 10 $\mu$m resolution. Specific heat was measured on warming procedure by the heat pulse relaxation method in PPMS-9.

X-ray photoemission spectroscopy (XPS) measurement was carried out in an ultrahigh-vacuum (UHV) system with $3\times10^{-10}$ Torr base pressure, equipped with a SPECS Phoibos 100 spectrometer and a non-monochromatized Al-K$_\alpha$ X-ray source ($h\nu$ = 1486.6 eV). The XPS peak positions were calibrated using adventitious C $1s$ at 284.8 eV. Single selected point unpolarized Raman spectrum experiment was performed using WITec confocal Raman microscope alpha 300 equipped with an red laser ($\lambda$ = 633 nm), an electron multiplying CCD detector and an 100$\times$/0.9NA objective lens. The Raman scattered light was focused onto a multi-mode fiber and monochromator with a 1800 line/mm grating. Samples for the room temperature XPS were exfoliated in air then be sputtered in UHV by 2$\times$10$^{-5}$ Torr of Ar$^+$ ions with kinetic energy of 2500 eV for 60 min in order to remove surface oxygen contamination. Samples for Raman experiments were freshly exfoliated.

First-principle calculations were performed using the Vienna ab initio simulation package (VASP) by relaxing the atomic positions simultaneously with changing of the unit cell volume and shape \cite{KresseG}. The energy convergence criterion was 10$^{-6}$ eV, while the interatomic forces were less than 0.02 eV/$\textrm{{\AA}}$. Energy cutoff for plane waves was 480 eV, the $k$-point network was 6$\times$9$\times$4 with the PBEsol as an exchange-correlation functional \cite{PerdewJ}. After lattice relaxation, the spin-orbit coupling (SOC) was included for both the density of states (DOS) and the band structure calculations. The band unfolding and the DOS figures were obtained by the Vaspkit program \cite{VASPKIT}. The high-symmetry points of the Brillouin zone were chosen on the same manner as in Ref. \cite{Ganose}

\section{RESULTS AND DISCUSSIONS}

The crystal structure of ZrTe$_3$ [Fig. 1(a)] consists of ZrTe$_3$ trigonal prisms building blocks stacked along the $b$-axis and forming an infinite quasi-1D chain \cite{Bra,Fur}. The monoclinic unit cell contains two neighboring chains bound via the nearest inter-chain Zr-Te1 to form layers in the $ab$-plane. On the other hand, Te2-Te3 atoms form a rectangular network with alternate distances of 2.79 and 3.10 {\AA} along the $a$-axis, and single distance of 3.93 {\AA} along the $b$-axis. The 5\% Cu,Ni intercalations induce expansion of the $a$- and $c$-axis lattice parameters, $\sim$ 0.2-0.3\% for Cu$_{0.05}$ZrTe$_3$ and 0.6-0.7\% for Ni$_{0.05}$ZrTe$_3$, respectively \cite{HC,XD1}. However the $b$-axis lattice constant is nearly unchanged. Figure 1(b) shows the normalized Zr, Cu and Ni $K$-edge X-ray absorption near edge spectroscopy (XANES) spectra measured at room temperature (RT). The Zr $K$-edge absorption energy at $\sim$ 18.008 keV for ZrTe$_3$ indicates a dominant Zr$^{4+}$ state \cite{Zahir}; it shifts to lower energies with Cu,Ni intercalation towards the Zr$^{2+}$ state. A weak pre-edge feature was observed in the Cu,Ni spectra, which is absent in that of Zr, along with the similar shape of Cu,Ni XANES, confirms that Cu,Ni atoms are well hybridized and intercalated in the vdW gap rather than substituted at Zr sites \cite{Westre}.

A clear resistivity upturn was observed at $T_{\textrm{CDW}}$ = 63 K for ZrTe$_3$, defined by the minima point in $d\rho/dT$, with the current parallel to the $a$-axis [Fig. 1(c)]. $T_{\textrm{CDW}}$ is suppressed to 58 and 41 K for Cu$_{0.05}$ZrTe$_3$ and Ni$_{0.05}$ZrTe$_3$, respectively, with a reduced amplitude. The residual resistivity ratio (RRR) of $\sim$ 12.7 for ZrTe$_3$ also decreases to 5.3 and 2.4 for Cu$_{0.05}$ZrTe$_3$ and Ni$_{0.05}$ZrTe$_3$, respectively, caused by increased disorder scattering. An abrupt resistivity drop is clearly seen, signaling the onset of SC, with zero resistivity at $T_\textrm{c}$ $\thickapprox$ 3.6 and 3.1 K for Cu$_{0.05}$ZrTe$_3$ and Ni$_{0.05}$ZrTe$_3$, respectively. For the current parallel to the $b$-axis, no CDW anomaly was observed but SC remains for (Cu,Ni)$_{0.05}$ZrTe$_3$; consistent with refs.\cite{HC,XD1}.

Figure 2(a) shows the temperature-dependent thermal conductivity $\kappa$(T) with the heat pulse applied in the $ab$-plane. The $\kappa$(300 K) shows a relatively low value of 1.63 W/K$\cdot$m for ZrTe$_3$, caused by the combination of low crystal symmetry and chemical composition with heavy elements. A clear kink in $\kappa$(T) is observed around $T_{\textrm{CDW}}$ for ZrTe$_3$, indicating strong e-p coupling. The value of $\kappa$(300 K) increases to $\sim$ 1.87 and 1.77 W/K$\cdot$m for Cu$_{0.05}$ZrTe$_3$ and Ni$_{0.05}$ZrTe$_3$, respectively. In addition, $\kappa$(T) is weakly temperature-dependent at high temperatures; the absence of a commonly observed maximum in $\kappa$(T) indicates a significant acoustic phonon scattering \cite{Bar}.

\begin{figure}
\centerline{\includegraphics[scale=1]{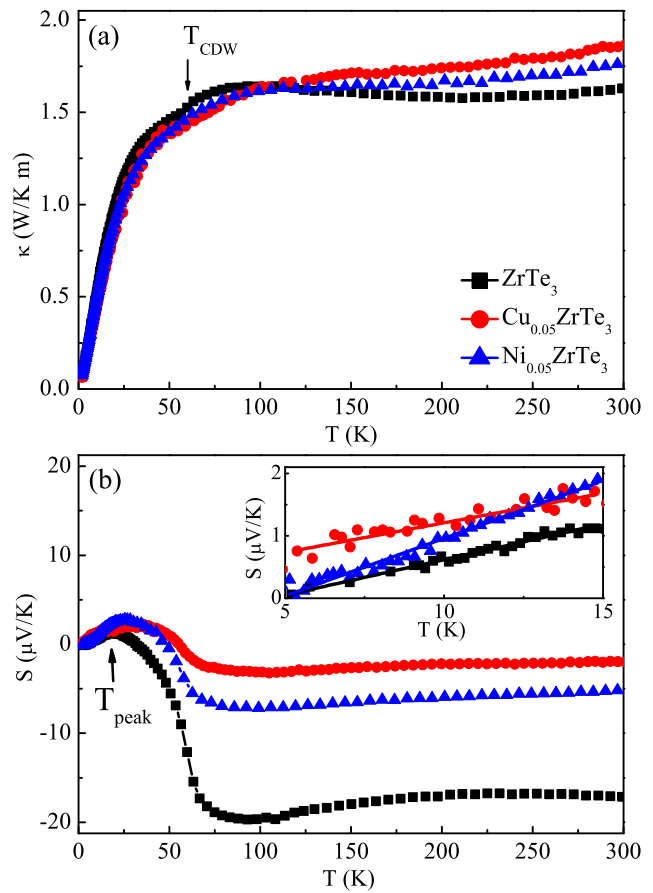}}
\caption{(Color online) Temperature-dependent (a) thermal conductivity $\kappa$(T) and (b) thermopower $S$(T) for indicated samples. Inset in (b) shows linear fit of $S$(T) from 5 to 15 K.}
\label{MTH}
\end{figure}

\begin{table*}
\caption{\label{tab} A set of parameters derived from the thermopower $S$(T) and specific heat $C_p$(T) for indicated single crystals.}
\begin{ruledtabular}
\begin{tabular}{llllllllllll}
  Samples & $S/T$($\mu$V/K$^2$) & $\gamma$(mJ/mol$\cdot$K$^2$) & $q$ & $T_\textrm{c}$(K) & $T_\textrm{F}$($10^3$K) & $T_\textrm{c}$/$T_\textrm{F}$($\times10^{-3}$) & $m^\ast/m_e$ & $\Theta_\textrm{D}$(K) & $T_\textrm{peak}$(K) & $T_{\textrm{CDW}}$(K)\\
  \hline
  ZrTe$_3$            & 0.114(3) & & & 2 & 3.7(1) & 0.54(1) & & & 19(3) & 60(2) \\
  Cu$_{0.05}$ZrTe$_3$ & 0.09(1) & 2.6(1) & 3.3(2) & 3.6 & 4.7(5) & 0.77(8) & 2.2(1) & 186(1) & 30(5) & 55(2) \\
  Ni$_{0.05}$ZrTe$_3$ & 0.188(4) & 2.7(1) & 6.7(1) & 3.1 & 2.3(1) & 1.35(6) & 2.8(1) & 192.4(1) & 26(3) & 58(2) \\
\end{tabular}
\end{ruledtabular}
\end{table*}

Figure 2(b) presents the temperature-dependent thermopower $S$(T) for ZrTe$_3$ and (Cu,Ni)$_{0.05}$ZrTe$_3$ crystals. Above 100 K, there are weak changes in $S$(T). With decreasing temperature, the $S$(T) of ZrTe$_3$ changes its slope at $T_{\textrm{CDW}}$, reflecting a FS reconstruction across the CDW transition. Then $S(T)$ changes its sign from negative to positive inside the CDW state featuring a peak value at $\sim$ 19(3) K, indicating a crossover behavior from dominant electron-to-hole-like conduction. Similar behavior is observed in (Cu,Ni)$_{0.05}$ZrTe$_3$ but with reduced absolute values of $S$ at high temperatures, showing an increase of the hole band contribution in (Cu,Ni)$_{0.05}$ZrTe$_3$. It can be seen that $T_{\textrm{CDW}}$ is suppressed to lower temperatures for (Cu,Ni)$_{0.05}$ZrTe$_3$ when compared to ZrTe$_3$, in line with the resistivity data [Fig. 1(c)]. It is plausible to consider both the electronic diffusion $S_{\textrm{diff}}$ and phonon drag $S_{\textrm{drag}}$ contribution to thermopower in ZrTe$_3$. The $S_{\textrm{drag}}$ term usually gives $\propto T^3$ for $T \ll \Theta_\textrm{D}$, $\propto 1/T$ for $T \gg \Theta_\textrm{D}$, and a peak feature at $(\frac{1}{4} \sim \frac{1}{5}) \Theta_\textrm{D}$. The peak feature in ZrTe$_3$ can not be simply attributed to only the phonon-drag effect since the peak temperature is lower than $\Theta_\textrm{D}/5 \approx$ 36.6(1) K; this is also applicable for (Cu,Ni)$_{0.05}$ZrTe$_3$ [see Table I]. However, the phonon drag should diminish by $1/T$ at high temperatures which is not seen here, pointing to the presence of diffusion contribution as well. At low temperatures, the diffusive Seebeck response of Fermi liquid dominates and is expected to be linear in $T$. In a single-band system, $S$(T) is given by \cite{Barnard,Miyake},
\begin{equation}
\frac{S}{T} = \pm \frac{\pi^2}{2}\frac{k_\textrm{B}}{e}\frac{1}{T_\textrm{F}} = \pm\frac{\pi^2}{3}\frac{k_\textrm{B}^2}{e}\frac{N(\varepsilon_\textrm{F})}{n},
\end{equation}
where $k_\textrm{B}$ is the Boltzmann constant, $T_\textrm{F}$ is the Fermi temperature related to the Fermi energy $\varepsilon_\textrm{F}$ and the density of states $N(\varepsilon_\textrm{F})$ as $N(\varepsilon_\textrm{F})$ = $3n/2\varepsilon_\textrm{F}$ = $3n/k_\textrm{B}T_\textrm{F}$, and $n$ is the carrier concentration. In a multiband system it gives the upper limit of $T_\textrm{F}$ of the dominant band. The derived $S/T$ from 5 to 15 K [inset in Fig. 2(c)] is $\sim$ 0.114(3) $\mu$V/K$^2$ for ZrTe$_3$; $\sim$ 0.09(1) and 0.188(4) $\mu$V/K$^2$ for Cu$_{0.05}$ZrTe$_3$ and Ni$_{0.05}$ZrTe$_3$, respectively. Then we calculate $T_\textrm{F}$ $\approx$ $3.7(1)\times10^3$ K for ZrTe$_3$; $4.7(5)\times10^3$ and $2.3(1)\times10^3$ K for Cu$_{0.05}$ZrTe$_3$ and Ni$_{0.05}$ZrTe$_3$, respectively. The ratio of $T_\textrm{c}/T_\textrm{F}$ characterizes the correlation strength in superconductors; $T_\textrm{c}/T_\textrm{F}$ is $\sim$ 0.1 in Fe$_{1+y}$Te$_{1-x}$Se$_x$, pointing to the importance of electron correlation; while $\sim$ 0.02 in a BCS superconductor LuNi$_2$B$_2$C \cite{Pourret}. Here the $T_\textrm{c}/T_\textrm{F}$ is $\sim$ 0.54(1)$\times10^{-3}$ for ZrTe$_3$; $\sim$ 0.77(8)$\times10^{-3}$ and 1.35(6)$\times10^{-3}$ for Cu$_{0.05}$ZrTe$_3$ and Ni$_{0.05}$ZrTe$_3$, respectively, indicating weak electron correlation strength but also a substantial enhancement in (Cu,Ni)$_{0.05}$ZrTe$_3$ when compared to ZrTe$_3$ [Table I].

The electronic specific heat can also be expressed as:
\begin{equation}
\gamma = \frac{\pi^2}{2}k_\textrm{B}\frac{n}{T_\textrm{F}} = \frac{\pi^2}{3}k_\textrm{B}^2N(\varepsilon_\textrm{F}).
\end{equation}
Combining equations (1) and (2) yields: $S/T = \pm \gamma/ne$; the units are V/K for $S$, J/K$^2$$\cdot$m$^3$ for $\gamma$, and m$^{-3}$ for $n$, respectively \cite{Behnia,PtSbSn}. In order to compare different materials, it is common to express $\gamma $ in J/mol$\cdot$K$^2$. Then we can define a dimensionless quantity $q=\frac{S}{T}\frac{N_\textrm{A}e}{\gamma}$, where $N_\textrm{A}$ is the Avogadro number; the calculated $q$ are 3.3(2) and 6.7(1) for Cu$_{0.05}$ZrTe$_3$ and Ni$_{0.05}$ZrTe$_3$, respectively. Given the volume of unit cell $\sim$ 0.23 nm$^3$, we obtain the carrier density per volume $n \approx 1.3(1) \times 10^{21}$ cm$^{-3}$ and $k_\textrm{F} \approx 3.4(1)$ nm$^{-1}$ for Cu$_{0.05}$ZrTe$_3$; $n \approx 6.5(1) \times 10^{20}$ cm$^{-3}$ and $k_\textrm{F} \approx 2.68(1)$ nm$^{-1}$ for Ni$_{0.05}$ZrTe$_3$. The effective mass $m^*$, derived from $k_\textrm{B}T_\textrm{F} = \hbar^2 k_\textrm{F}^2/2m^*$, are 2.2(1) and 2.8(1) $m_e$ for Cu$_{0.05}$ZrTe$_3$ and Ni$_{0.05}$ZrTe$_3$, respectively. Both the ratio of $T_\textrm{c}$ to the Fermi temperature $T_\textrm{F}$ and the effective mass $m^*$ rise in (Cu,Ni)$_{0.05}$ZrTe$_3$ when compared to ZrTe$_{2.96}$Se$_{0.04}$ [$T_\textrm{c}$/$T_\textrm{F}$ = 0.0016(1) and $m^\ast$ = 1.53(1) $m_e$] \cite{APL}, indicating the increase of correlation strength with Cu,Ni-intercalation when compared to Se-substitution.

\begin{figure}
\centerline{\includegraphics[scale=1]{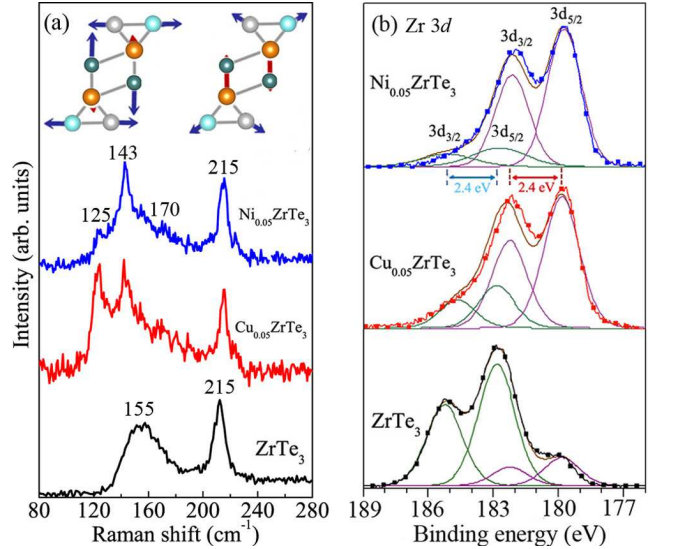}}
\caption{(Color online) (a) Raman peaks and (b) Zr $3d$ XPS spectra measured at room temperature for indicated samples.}
\label{MTH}
\end{figure}

Figure 3(a) shows the Raman peaks for ZrTe$_3$ ($\sim$ 215 and 155 cm$^{-1}$) \cite{Yang,APL}, corresponding to $A_g$ modes which involve atomic movements in the $ac$-plane \cite{Hu}. The mode at 155 cm$^{-1}$ hosts momenta that have a larger component along the $a$-axis of Te2-Te3 chains. The conductive electrons belong to the quasi-1D FS are mainly contributed by the Te 5$p_x$ orbitals of Te2-Te3 chains, which can be only scattered forward or backward along Te2-Te3 chains. Consequently, phonons on this mode will experience stronger scattering with conduction electrons of quasi-1D FS, causing the wide peak as observed due to strong e-p coupling by satisfying $\emph{\textbf{q}}\cdot \emph{\textbf{v}}_\textrm{F} \approx \omega_{q\nu}$, where $\emph{\textbf{v}}_\textrm{F}$ is Fermi velocity and $\omega_{q\nu}$ is the phonon energy \cite{Hu}. The mode at 215 cm$^{-1}$ also involves the Te2-Te3 chains, but does not exhibit a large linewidth as that at 155 cm$^{-1}$. This can be explained by the fact that this mode is not coupled with the quasi-1D conduction electrons and can only be affected by non-conductive 3D-like electrons \cite{Hu}. The sharpness of this peak suggest the good crystallinity of ZrTe$_3$.

For (Cu,Ni)$_{0.05}$ZrTe$_3$, the mode at 215 cm$^{-1}$ is almost unaffected; in contrast, the mode at 155 cm$^{-1}$ are significantly modulated. The position of highest peak red shift $\sim$ 12 cm$^{-1}$ from 155 to 143 cm$^{-1}$, suggesting tensile lattice deformation of Te2-Te3 chains along the $a$- or $c$-axis. This is quantitatively consistent with the structure analysis that Cu,Ni intercalations induce expansion of the $a$- and $c$-axis lattice parameters. In addition, broad peaks at $\sim$ 125 and 170 cm$^{-1}$ are resolved for (Cu,Ni)$_{0.05}$ZrTe$_3$. Weak Raman peaks at $\sim$ 122 and 177 cm$^{-1}$ that also belong to the eight $A_g$ modes in the $ac$-plane were observed for pure ZrTe$_3$ at 70 K with a $cc$ polarized laser \cite{Hu}. Therefore, the lattice modulation caused by Cu,Ni intercalations enabled these vibration modes become observable at RT here with an unpolarized laser. More importantly, the major peak at 143 cm$^{-1}$ with narrow linewidth is much sharper comparing to the corresponding peak for ZrTe$_3$ at 155 cm$^{-1}$, which suggests the decoupling of quasi-1D conduction electrons with phonons at RT. This probably due to the e-p coupling condition of $\emph{\textbf{q}}\cdot \emph{\textbf{v}}_\textrm{F} \approx \omega_{q\nu}$ being less satisfied since the phonon energy is reduced as consequence of intercalation-induced lattice expansion. Thus, the Raman results proved that the suppressed CDW is structural distortion-driven suppression of the quasi-1D conductive e-p coupling. Structurally driven CDW suppression due to hydrostatic pressure or Se doping-induced long range disorder of lattices in the $c$- or $a$-axis, respectively, were also reported \cite{Gleason,APL}. In these two cases, the suppression of e-p coupling were evidenced by the reduced Raman intensity ratio of the broad peak at 155 cm$^{-1}$ to the peak at 215 cm$^{-1}$; structurally, the lattice parameters in the $c$-axis or the size of crystal unit cell were compressed. While in (Cu,Ni)$_{0.05}$ZrTe$_3$, in contrast, the suppression of e-p coupling is evidenced by the reduced-linewidth of the major peak at 155 cm$^{-1}$; structurally, the lattice parameters in the $a$- and $c$-axis are expanded, indicating that CDW suppression can be realized through structural modulation in various pathways.

The XPS shows that both ZrTe$_3$ and (Cu,Ni)$_{0.05}$ZrTe$_3$ have mixed chemical states of Zr$^{4+}$ and Zr$^{2+}$ as evidenced by two sets of binding energies of Zr $3d_{5/2}$ and Zr $3d_{3/2}$ at $\sim$ 183, 185.4 eV, and 180, 182.4 eV [Fig. 3(b)]. However, the atomic concentration ratio of Zr$^{4+}$ to Zr$^{2+}$ are dramatically altered by the intercalation, which are 4.3, 0.32 and 0.23, for ZrTe$_3$, Cu$_{0.05}$ZrTe$_3$ and Ni$_{0.05}$ZrTe$_3$, respectively. The average valence Zr is $\sim$ +3.6 for ZrTe$_3$, which decreases to $\sim$ +2.5 for Cu$_{0.05}$ZrTe$_3$ and +2.4 for Ni$_{0.05}$ZrTe$_3$, respectively, in agreement with the trend in the XANES measurement. The Zr-Te bonds around Zr are spatially anisotropic in bond type and spatial distribution, causing spatially anisotropic charge orbital environment around Zr and consequently inhomogeneous Zr valence. The XANES and Raman results suggested that the lattice parameters of ZrTe$_3$ are changed by Cu,Ni intercalations, thus the spatial anisotropic degree of Zr-Te bonds around Zr for (Cu,Ni)$_{0.05}$ZrTe$_3$ can be also expected to be changed, resulting in the observed change in ratio of Zr$^{4+}$ to Zr$^{2+}$. Compared to Hf,Se-doping-caused unit cell reduction accompanying by decreased Zr$^{2+}$ to Zr$^{4+}$ ratio \cite{Gleason,APL}, Cu,Ni-intercalations-induced unit cell expansion is accompanied by increased Zr$^{2+}$ to Zr$^{4+}$ ratio. Since 5\% Ni induced a more expanded unit cell than Cu, Zr$^{2+}$ to Zr$^{4+}$ ratio also increased more. The hole-type 3D FS at the BZ center is mainly contributed by Zr $d$ orbital. ARPES shows that the 3D FS is slightly reduced in ZrTe$_{2.96}$Se$_{0.04}$ when compared to ZrTe$_3$ \cite{4}. The reduced Zr$^{2+}$ to Zr$^{4+}$ ratio in ZrTe$_{2.96}$Se$_{0.04}$ \cite{APL} is coincident with the reduction of hole-type carrier numbers in the 3D band as confirmed by ARPES \cite{4}, thus it is possible that Zr$^{2+}$ contributed to the 3D FS though it is not much involved in the formation of CDW. The quasi-1D FS responsible for CDW and SC is mainly distributed on the BZ edge, in particular in the $D$-pocket \cite{4}, is more sensitive to the change in BZ size modulated by lattice expansion/reduction.

\begin{figure}
\centerline{\includegraphics[scale=0.45]{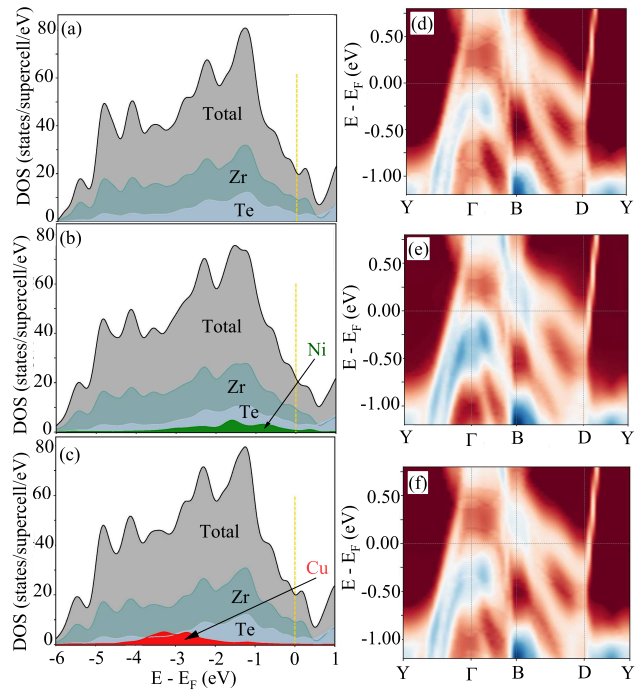}}
\caption{(Color online)Total and atomic contribution to density of states (a-c) and calculated band structures near $\Gamma$, B, D, and Y high symmetry points (d-f) for ZrTe$_3$, Cu$_{0.06}$ZrTe$_3$ and Ni$_{0.06}$ZrTe$_3$.}
\label{MTH}
\end{figure}

For insight into the impact of Cu, Ni intercalation we further performed comparative first-principles calculations of the density of states (DOS) [Fig. 4(a-c)] and band structure [Fig. 4(d-f)]. An intercalated atom in ZrTe$_3$, either Cu or Ni, resides in a pentacoordinated Wyckoff 2$e$ position within the vdW space gap. This position is in agreement with the most favorable interstitial position \cite{Ganose} as it is energetically favored when compared to tetrahedrally coordinated 2$a$ and 2$b$ positions. Calculated DOS for ZrTe$_3$ [Fig. 4(a)] is in agreement with previous \cite{KuboY}. Upon intercalation, Ni contribution to DOS at the Fermi level is somewhat higher when compared to Cu [Fig. 4(b,c)], in agreement with the effective mass $m^*$ (Table I). The effects of intercalation in vdW gap of ZrTe$_3$ are not substantial at the Fermi level. They are mainly visible in bands along $B$-$D$ high symmetry direction as the increase in the spectral weight of the bands at energies (0.2-0.25) eV above the Fermi level whereas quasi-1D band dispersion is unchanged between $D$ and $Y$ indicating negligible charge transfer from intercalated atoms to Te2-Te3 chains. On the other hand, we notice a shift near $\Gamma$ for intercalated crystals [Fig. 4(e,f)] for the Zr 4$d$-derived band at about -0.25 eV; the shift has been associated with Zr-Ni charge transfer in Ni$_x$ZrTe$_3$ \cite{Ganose}.

For doped ZrTe$_3$, the electronic structure of quasi-1D FS i.e., the FS nesting condition is modulated via atomic structural modulation that are not spatial uniformly through out the entire ZrTe$_3$ crystal. In other words, the lattice/BZ modulations are more likely in short range order, consistent with the observation of reduced intensity or linewidth of the Raman peak at 155 to 143 cm$^{-1}$ [Fig. 3(a)]. Long range static CDW appears to compete with SC \cite{Ghi,EH,TP}, while short range CDW correlations are present over a substantial range of carrier concentrations \cite{Ghi,DL} that support the appearance of SC. This can explain the higher $T_c$ for Se-doped or Cu,Ni-intercalated ZrTe$_3$ than that of pristine ZrTe$_3$. For Hf-doped ZrTe$_3$, because most of Te2-Te3 chains are relaxed, thus are kept the original lattice periodicity, the long range static CDW is preserved and is more stable at the cost of reduced density of states of quasi-1D band at the Fermi level that are necessary for SC formation, therefore, rendering suppressed SC. Compared with TiSe$_2$ \cite{Morosan} and other CDW materials, the SC can be substantially enhanced by Cu,Ni-intercalation in ZrTe$_3$ while the CDW is only slightly suppressed. Recently, the thermal destruction of CDW in ZrTe$_3$ studied by scanning tunneling microscopy (STM) reveals a weak to strong impurity pinning across the CDW transition \cite{Jiaxin}, calling for further STM study on the effects of Hf,Se-doping and Cu,Ni-intercalation on the collective CDW and SC orders in ZrTe$_3$.

\section{CONCLUSIONS}

In summary, 5\% Cu,Ni intercalations in the vdW gap of ZrTe$_3$ result in increase in $T_c$ as well as electron correlation strength. The ratio of Zr$^{2+}$ to Zr$^{4+}$ atoms also increases, resulting in band shifts near the $\Gamma$ point in the Brillouen zone. Intercalation results in a tensile lattice deformation of Te2-Te3 atomic chains along the $a$-/$c$-axis, and consequent expansion of the unit cell along these directions. This reduces the e-p coupling of quasi-1D FS due to reduction of phonon energy but also affects its nesting condition via atomic structural modulation, thus revealing the mechanism of CDW suppression. It exposes the important role of mixed valence in the competition of collective orders for the ground state at the Fermi surface.

\section*{Acknowledgements}

We thank Hengxin Tan and Binghai Yan for useful communication. Work at Brookhaven National Laboratory (BNL) is supported by the Office of Basic Energy Sciences, Materials Sciences and Engineering Division, U.S. Department of Energy under Contract No. DE-SC0012704. This research used the 8-ID beamline of the National Synchrotron Light Source II, and resources of the Center for Functional Nanomaterials, which is a U.S. Department of Energy Office of Science User Facility, at BNL under Contract No. DE-SC0012704. First principle calculations were supported by the Ministry of Education, Science, and Technological Development of the Republic of Serbia.

$^{\S}$Present address: Los Alamos National Laboratory, Los Alamos, New Mexico 87545, USA. $^{*}$Present address: Anhui Province Key Laboratory of Condensed Matter Physics at Extreme Conditions, High Magnetic Field Laboratory, Chinese Academy of Sciences, Hefei 230031, China. $^{\dag}$Present address: Department of Physics and Beijing Key Laboratory of Opto-electronic Functional Materials $\&$ Micro-nano Devices, Renmin University of China, Beijing 100872, China. $^{\ddag}$Present address: ALBA Synchrotron Light Source, Cerdanyola del Valles, E-08290 Barcelona, Spain.

$^{\P}$yuliu@lanl.gov
$^{\|}$petrovic@bnl.gov

\end{document}